\documentclass[fleqn,usenatbib]{mnras}

\usepackage{newtxtext,newtxmath}

\usepackage{aas_macros}
\usepackage{orcidlink}
\usepackage{fontawesome}
\usepackage{float}
\usepackage{amssymb}

\usepackage[T1]{fontenc}

\DeclareRobustCommand{\VAN}[3]{#2}
\let\VANthebibliography\thebibliography
\def\thebibliography{\DeclareRobustCommand{\VAN}[3]{##3}\VANthebibliography}

\usepackage{graphicx}	
\usepackage{amsmath}	
\usepackage{amssymb}	
\usepackage{caption}
\usepackage{subcaption}
\usepackage{graphicx}
\usepackage{subcaption}
\usepackage{multirow}
\usepackage{array}
\usepackage{comment}
\usepackage{xcolor}
\newcolumntype{P}[1]{>{\centering\arraybackslash}p{#1}}

\newcommand{\be}{\begin{equation}}
\newcommand{\ee}{\end{equation}}
\newcommand{\bea}{\begin{eqnarray}}
\newcommand{\eea}{\end{eqnarray}}

\newcommand{\bv}{{\bf{v}}}
\newcommand{\bn}{{\bf{n}}}
\newcommand{\nn}{\nonumber}

\title[Cosmic dipole detection with ET and CE]{Detection and estimation of the cosmic dipole with the Einstein Telescope and Cosmic Explorer}

\author[Mastrogiovanni et al.]{S.~Mastrogiovanni$^{1,2}$\thanks{smastro@oca.eu}\orcidlink{0000-0003-1606-4183},
 C.~Bonvin$^{3}$\orcidlink{0000-0002-5318-4064}, G.~Cusin$^{3,4}$,
S.~Foffa$^{3}$\orcidlink{0000-0002-4530-3051}
\\
$^{1}$ INFN, Sezione di Roma, I-00185 Roma, Italy\\
$^{2}$ Artemis, Université Côte d’Azur, Observatoire de la Côte d’Azur, CNRS, F-06304 Nice, France\\
$^{3}$Universit\'e de Gen\`eve, D\'epartement de Physique Th\'eorique and Gravitational Wave Science Center, 24 quai Ernest-Ansermet, \\CH-1211 Gen\`eve 4, Switzerland\\
$^{4}$Sorbonne Université, CNRS, UMR 7095, Institut d'Astrophysique de Paris, 75014 Paris, France \\
}
\date{Accepted XXX. Received YYY; in original form ZZZ\\
ET-0195A-22}
\pubyear{2022}

\begin{document}
\label{firstpage}
\pagerange{\pageref{firstpage}--\pageref{lastpage}}
\maketitle

\begin{abstract}
One of the open issues of the standard cosmological model is the value of the cosmic dipole measured from the Cosmic Microwave Background (CMB), as well as from the number count of quasars and radio sources. These measurements are currently in tension, with the number count dipole being 2-5 times larger than expected from CMB measurements. This discrepancy has been pointed out as a possible indication that the cosmological principle is not valid. 
In this paper, we explore the possibility of detecting and estimating the cosmic dipole with gravitational waves (GWs) from compact binary mergers detected by the future next-generation detectors Einstein Telescope and Cosmic Explorer. We model the expected signal and show that for binary black holes, the dipole amplitude in the number count of detections is independent of the characteristics of the population and provides a systematic-free tool to estimate the observer velocity. We introduce techniques to detect the cosmic dipole from number counting of GW detections and estimate its significance.  We show that a GW dipole consistent with the amplitude of the dipole in radio galaxies would be detectable with $>3\sigma$ significance with a few years of observation ($10^6$ GW detections) and estimated with a $16\%$ precision, while a GW dipole consistent with the CMB one would require at least $10^7$ GW events for a confident detection. We also demonstrate that a total number $N_{\rm tot}$ of GW detections would be able to detect a dipole with amplitude $v_o/c \simeq1/\sqrt{N_{\rm tot}}$.
\end{abstract}




\section{Introduction}

Our motion through the Universe generates a dipole in both the temperature anisotropies of the Cosmic Microwave Background (CMB) \citep{2020A&A...641A...6P} and in the angular distribution of electromagnetic sources \citep{2017MNRAS.471.1045C,2018JCAP...04..031B,2021ApJ...908L..51S,2021A&A...653A...9S,Secrest:2022uvx}. If the cosmological principle is valid, these two measures should have consistent values. However, it is a longstanding problem that number counts of radio sources and of quasars at low and intermediate redshifts exhibit a dipole that is well aligned with that of the CMB but with an amplitude which is 2-5 times larger than expected, leading to a tension reaching up to $\sim 5 \sigma$. In \citet{Dalang:2021ruy} it is argued that this tension might be alleviated once one takes into account the redshift evolution of the population of sources, and the value that the evolution rate should have in order to remove the tension is found. 

Gravitational-wave (GW) sources observed at cosmological distances can shed light on the cosmic dipole problem. Since the first historical detection of~\citet{LIGOScientific:2016aoc} during the first scientific observation run (O1), the rate of observed GW events has drastically increased, to reach roughly one detection per week in the O3 run. In total, about 90 binary black hole (BBH) coalescences, as well as two binary neutron star (BNS) and two neutron star - black hole mergers have been detected so far~\citep{2021arXiv211103606T}. Thanks to this abundance of new data, which will further accumulate in the forthcoming future \citep{2018LRR....21....3A}, it is possible to probe cosmology \citep{2021ApJ...909..218A,https://doi.org/10.48550/arxiv.2111.03604,2022PhRvD.105f4030M,2022arXiv220200025L} and astrophysical rates of compact binary coalescences \citep{2019ApJ...882L..24A,2021ApJ...913L...7A,2021arXiv211103634T} using GW observations.  

It is thus not surprising that recently, it has been proposed to use GW sources as a new and independent probe to measure the cosmic dipole. A first approach proposed to measure the cosmic dipole is by studying anisotropies in the GW stochastic background  \citep{Cusin:2022cbb,  DallArmi:2022wnq, LISACosmologyWorkingGroup:2022kbp}. However, stochastic GW backgrounds have not been detected yet.
Another possibility, is to study the sky distribution of the transient GW sources currently detected \citep{2018PhRvD..97j3005C,2022arXiv220407472K,2022arXiv220705792E}. Although GW sources are the central paradigm of all these works, the methods employed significantly differ. In \citet{2018PhRvD..97j3005C}, the authors perform a forecast on the accuracy with which the cosmic dipole would be detected by fitting for the (modified) luminosity distance distribution of GW events. In \citet{2022arXiv220407472K}, the authors try instead to constrain the dipole anisotropy using the BBHs mass distribution from current GW events \citep{2021arXiv211103606T} finding that the mean mass is higher in the direction of the CMB dipole and invoking the need of further study to understand the origin of their findings. In contrast to \citet{2021MNRAS.501..970S}, \citet{2022arXiv220407472K}, \citet{2022arXiv220705792E} find no-evidence of anisotropies, using 63 GW sources distributed over the sky, and implementing a hierachical Bayesian analysis that takes into account selection biases.

In our work, we focus on the detection and estimation of the cosmic dipole with Einstein Telescope (ET) \citep{2010CQGra..27s4002P} and Cosmic Explorer (CE) \citep{Reitze:2019iox}.
These detectors, along with LISA \citep{2017arXiv170200786A},  belong to the so-called next generation (XG) of GW detectors and represent the aim of the community to substantially scale-up the experience of the LIGO-Virgo era. In this paper, we show that the number counts of BBH sources will likely offer an optimal tool to detect the cosmic dipole. We build an estimator, that has the advantage of being independent of unknown characteristics of the GW sources (or their evolution) and that will therefore allow us to infer our motion with respect to the cosmological frame in a robust way.

This paper is structured as follows. In Sec.~\ref{sec:2}, we derive the dipole modulation in the GW sources number counts, due to the observer velocity with respect to the cosmological frame. 
We show in this section that GW number counting of BBHs offers a clean tool to evaluate anisotropies.
In Sec.~\ref{sec:3}, we simulate GW detections using a detector network composed by ET and two CEs (ET+2CE). We discuss detection capabilities and the measurement process using a frequentist and Bayesian approach.  In Sec.~\ref{sec:4}, we discuss possible limiting factors of our approach in light of the possibility of constraining the cosmic dipole with GW sources. Finally, in Sec.~\ref{sec:5} we draw our conclusions.





\section{Modeling the impact of the observer velocity on GW number counts}
\label{sec:2}

We write down explicitly the expression for the cosmic dipole in the GW source distribution. In Sec.~\ref{sec:2.1}, we derive the theoretical modeling for the dipole induced by the observer velocity in the number counts of GW signals    emitted by  compact binary coalescences. In Sec.~\ref{sec:2.3}, we define a statistical estimator for the cosmic dipole.

\subsection{Theoretical framework}
\label{sec:2.1}
We define the number of GW sources $N_{\rm det}$ detectable per unit solid angle $d\Omega$ and distance bin $dr$, in direction $\bn$ at comoving distance $r$ and with signal-to-noise ratio (SNR) $\rho$ larger than a given threshold $\rho_*$ as
\be\label{int}
\frac{d N_{\rm det}}{d\Omega dr}(r, \bn, \rho>\rho_*)\equiv \int_{\rho_*}^{\infty} d\rho \frac{d N}{d\Omega dr}(r, \bn, \rho)\,.
\ee 
This number depends on the direction of observation $\bn$ due to three effects. First, the GW sources are not perfectly homogeneously distributed: they live in galaxies which follow the large-scale structure of the Universe. Second, the propagation of GWs is affected by inhomogeneities along the trajectory, that change the  apparent distribution of sources over the sky. Line of sight effects include lensing and local matter effects, and also effects due to the source peculiar motion see e.g.~\cite{Takahashi:2003ix, Barausse:2014tra, Bonvin:2016qxr, Sberna:2022qbn, Cusin:2020ezb, Toubiana:2020drf, Cusin:2021rjp, Bonvin:2022mkw, Toscani:2023sfc} and \cite{Cusin:2017fwz, Pitrou:2019rjz} in the context of a stochastic background. Finally, the motion of the observer with respect to the source rest frame generates a further anisotropy in the observed distribution. In this work, we are interested in the latter effect, which gives a dipolar modulation in the distribution of sources. The other two effects will also have a dipolar contribution, but, as has been shown for quasars~\citep{2021ApJ...908L..51S}, this clustering dipole is expected to be negligible compared to the kinematic dipole. We checked that this hypothesis is valid also for the BBH and BNS merger rate model used in this work (see Sec.~\ref{sec:3}) for more details. 

The cosmic dipole averaged over all distances $r$ is obtained from the difference between Eq.~\eqref{int} and its angular average, integrated over $r$. Namely,  
\begin{align}
&\Delta(\bn)\equiv \mathcal{D} [\bn\cdot \hat{\bv}_o]\nn\\
&=\dfrac{\int dr \left[\frac{dN_{\rm det}}{d\Omega dr}(r, \bn, \rho>\rho_*)-\frac{d\bar{N}_{\rm det}}{d\Omega dr}(r, \rho>\rho_*)\right]}{\int dr \frac{d\bar{N}_{\rm det}}{d\Omega dr}(r, \rho>\rho_*)}\,,
\label{dipole0}
\end{align}
where $\bv_o$ denotes the observer velocity and  $\hat{\bv}_o\equiv \bv_o/|\bv_o|$ is its direction. The angular average over the sky, denoted with a bar, is given by 
\be
\frac{d\bar{N}_{\rm det}}{d\Omega dr}(r, \rho>\rho_*) \equiv \frac{1}{4\pi} \int d\Omega \frac{dN_{\rm det}}{d\Omega dr}(r, \bn, \rho>\rho_*)\,.
\ee

Eq.~\eqref{int} depends on the observer velocity through two effects. First, the observed solid angle is affected by aberration:
\be
d\Omega=d\bar{\Omega}\left(1-2 \bn\cdot \frac{\bv_o}{c}\right)\,.
\label{eq:domegabar}
\ee
And second, the SNR threshold at the observer, $\rho_*$, (which is a fixed number) corresponds to different emitted GW power for sources situated in different directions. Since we want to relate the $\rho_*$ dependence of Eq.~\eqref{int} to the astrophysical distribution of GW sources and the effect of the kinematic dipole, we factorize the SNR as follows
\begin{align}
\rho^2(r,\bn)=I\times P(r,\bn) \, .
\label{eq:model}
\end{align}
The part $I$ is an `intrinsic'' part that accounts for the fact that the SNR depends on the astrophysical properties of the sources (such as masses). The ``propagation'' part, $P(r,\bn)$ on the other hand describes how the SNR depends on the relative position of the source and the observer. Since distances and redshifts are affected by the observer velocity, this propagation part is not isotropic: it depends on $\bn$ (more precisely it depends on the angle between $\bn$ and the observer velocity). We will see later in this section how to define these quantities for GWs emitted from inspiralling binaries.

A source sitting at position $(r,\bn)$ with an SNR above threshold, $\rho(r,\bn)>\rho_*$, must have intrinsic properties $I>I_*(r,\bn)$ where
\begin{align}
 I_*(r,\bn)\equiv \frac{\rho_*^2}{P(r,\bn)}\, .  \label{eq:Istar} 
\end{align}
We see that because $P(r,\bn)$ depends on $\bn$, a fixed SNR threshold $\rho_*$ corresponds to different intrinsic properties $I_*(r,\bn)$ in different directions. Using that the observer velocity is small compared to the Hubble flow, we can then Taylor expand Eq.~\eqref{int} around the homogeneous background. We obtain
\begin{align}
\frac{d N_{\rm det}}{d\Omega dr}(r, \bn, \rho>\rho_*)
&=\frac{d N_{\rm det}}{d\Omega dr}(r, \bn, I>I_*(r,\bn))\nn\\
&\simeq\frac{d N_{\rm det}}{d\bar{\Omega }dr}(r, I>\bar{I}_*(r))\left(1+2\bn\cdot \frac{\bv_o}{c}\right)\nn\\
&+\frac{\partial }{\partial I_*}\left(\frac{d N_{\rm det}}{d\Omega dr}(r, I>I_*)\right)_{I_*=\bar{I}_*}\delta I_*(r,\bn)\,, \label{eq:Npert}
\end{align} 
where we have used Eq.~\eqref{eq:domegabar} in the second line.

To compute $\delta I_*(r,\bn)$ we need a model for the SNR. For a binary system of compact objects, the SNR at the zero Post-Newtonian (0PN) order is given by~\citep{Finn:1992xs}
\be
\rho^2(r,\bn, m, \mathcal{M})= \frac{5}{96\pi^{4/3}} \frac{\Theta^2}{D_L^2(r,\bn)} (G \mathcal{M}_z)^{5/3} \mathcal{F}(f^z_{\rm ISCO}(m))\,,
\label{eq:SNR1}
\ee
where $\mathcal{M}_z=\mathcal{M}(1+z)$ is the \textit{redshifted}  chirp mass of the system, $m=m_1+m_2$ is the total mass, $\Theta^2$ is a geometrical factor that depends on the inclination of the binary and on the antenna pattern of the detector, and $D_L$ denotes the luminosity distance. The $\mathcal{F}$ quantifies the sensitivity of the GW detector, namely 
\be\label{f731}
\mathcal{F}(f^z_{\rm ISCO}(m))\equiv \int_0^{2f^z_{\text{ISCO}}} df\left[f^{7/3} S_n(f)\right]^{-1}\,,
\ee
where $S_n(f)$ is the detector Power Spectral Density (PSD).
In Eq.~\eqref{f731} the upper integration bound is given by the redshifted  frequency corresponding to the innermost stable circular orbit ($\text{ISCO}$) of the system, i.e. the frequency at which we consider the inspiraling phase of the system to end in our approximation. It is defined as \citep{Maggiore:1900zz}
\be
f_{\text{ISCO}}\equiv\frac{1}{6\sqrt{6}(2\pi)}\frac{c^3}{Gm}\simeq  2.2. \text{kHz}\left(\frac{M_{\odot}}{m}\right)\,,
\label{eq:isco}
\ee
and $f^z_{\text{ISCO}}=f_{\text{ISCO}}/(1+z)$. 
Note that at this frequency the 0PN approximation in Eq.~\eqref{eq:SNR1} used for the SNR estimate becomes inaccurate.

Referring to Eq.~\eqref{eq:model}, we now define the intrinsic part of the SNR as the one that depends only on the intrinsic properties of the source
\begin{align}
I(\mathcal{M},m)\equiv \frac{5}{96\pi^{4/3}} \Theta^2 (G \mathcal{M})^{5/3} \mathcal{F}(f_{\rm ISCO}(m))\, ,    
\end{align}
and the propagation part as the part that depends on the relative position of the source and the observer, i.e.\ on the redshift and luminosity distance of the source
\begin{align}
P(r,\bn)\equiv \frac{(1+z(r,\bn))^{5/3}}{D_L^2(r,\bn)}\frac{\mathcal{F}(f^z_{\rm ISCO}(m,r,\bn))}{\mathcal{F}(f_{\rm ISCO}(m))}\, .   \label{eq:P}
\end{align}
Since the redshift and the luminosity distance are affected by the observer velocity, the propagation factor, $P$ depends directly on $\bv_0$. Note that the redshift perturbation also enters via the upper bound of the integral $f^z_{\rm ISCO}$, which reflects the fact that the observer velocity shifts the observed frequency of the ISCO. Inserting~\eqref{eq:P} into~\eqref{eq:Istar} and expanding at linear order in the velocity we obtain
\begin{align}
 \delta I_*(r,\bn)=\bar{I}_*(r)&\Bigg[2\frac{\delta D_L}{\bar{D}_L}-\frac{5}{3}\frac{\delta z}{1+\bar{z}}\\
 &+\frac{2f^{\bar{z}}_{\rm ISCO}}{\mathcal{F}(f^{\bar{z}}_{\rm ISCO})}
 \frac{\left(2f^{\bar{z}}_{\rm ISCO}\right)^{-7/3}}{
 S_n\left(2f^{\bar{z}}_{\rm ISCO}\right)}\frac{\delta z}{1+\bar{z}}\Bigg]  \, .\nn 
\end{align}
Using that the redshift pertubation and luminosity distance perturbations are given by
\be\label{eq:pert}
\frac{\delta z}{1+\bar{z}}=-\bn\cdot \frac{\bv_o}{c}\,,\quad \mbox{and}\quad
\frac{\delta D_L}{\bar{D}_L}=-\bn\cdot \frac{\bv_o}{c}\,,
\ee
we find
\begin{align}
 \delta I_*(r,\bn)=-\bar{I}_*(r)&\Bigg[\frac{1}{3}+\frac{2f^{\bar{z}}_{\rm ISCO}}{\mathcal{F}(f^{\bar{z}}_{\rm ISCO})}\frac{\left(2f^{\bar{z}}_{\rm ISCO}\right)^{-7/3}}
 {S_n\left(2f^{\bar{z}}_{\rm ISCO}\right)}\Bigg]\bn\cdot \frac{\bv_o}{c}  \, . \label{eq:Ipert}
 \end{align}

Finally, we need to compute the variation of the cumulative number of events above threshold:
\be
\frac{\partial }{\partial I_*}\left(\frac{d N_{\rm det}(r,I>I_*)}{d\Omega dr}\right)_{I_*=\bar{I}_*}\,.
\ee
Without loss of generality, we assume that GW events are distributed in a window of $I$, i.e.\ $I \in [I_{\text{min}}, I_{\text{max}}]$. The intrinsic part of the SNR, $I$, is indeed directly related to the chirp mass of the system, and to the total mass (through $f_{\rm ISCO}$). Both these quantities have a given distribution with finite width. Therefore only a range of values of $I$ are physical. We can write 
\begin{align}
\label{eq:defs}
\frac{\partial }{\partial I_*}\left(\frac{d N_{\rm det}(r,I>I_*)}{d\Omega dr}\right)_{I_*=\bar{I}_*}=&-\frac{s(I_*)}{I_*}\frac{d N_{\rm det}(r,I>I_*)}{d\Omega dr}\,,
\end{align} 
where the parameter $s$ is defined through
\begin{align}
 s(I_*)=s(\rho_*,r)\equiv -\frac{\partial\ln \left(\frac{d N_{\rm det}(r,I>I_*)}{d\Omega dr}\right)}{\partial\ln I_*}\, . 
 \label{eq:s}
 \end{align}
 This parameter directly depends on the population of sources. It is non-zero only if $I_{\rm min}<I_*<I_{\rm max}$, or similarly if $\rho_{\rm min}(r)<\rho_*<\rho_{\rm max}(r)$, i.e. when we are dealing with a population of sources with a significant fraction of events across threshold:
 \begin{align}
   s(\rho_*,r)\propto \Theta(\rho_*-\rho_{\text{min}})\Theta(\rho_{\text{max}}-\rho_*)\, .  
 \end{align}

Inserting Eqs.~\eqref{eq:defs},~\eqref{eq:Ipert} and~\eqref{eq:domegabar} into~\eqref{eq:Npert}, and subtracting the angular average, we find for the dipole 
\begin{align}
\mathcal{D}\left[{\bf n}\cdot{\bf v}_0\right]&=\bn\cdot \frac{\bv_o}{c}\int_0^{\infty} dr f(r)\left[2+s(r,\rho_*)\left(\frac{1}{3}+\mathcal{A}(r)\right)\right] \nn\\
&\equiv\alpha\,\bn\cdot \frac{\bv_o}{c} \,,
\label{eq:dipole}
\end{align}
where we implicitly defined the parameter $\alpha$ that will be used later on in the analysis. The function $f(r)$ denotes the radial distribution of sources 
\be\label{fr}
f(r)\equiv\frac{\frac{dN}{drd\Omega}}{\int_0^{\infty} dr \frac{dN}{d\Omega dr}}\,,
\ee
and $\mathcal{A}$ is given by the second term in Eq.~\eqref{eq:Ipert}, averaged over all sources at distance $r$: 
\begin{align}
\mathcal{A}(r) \equiv & \int dm_{1} dm_{2}\, p(m_{1}, m_{2})  \frac{1}{\mathcal{F}\left(\frac{f_{\text{ISCO}}}{(1+z)}\right)} \nn \\ 
& \times\left(\frac{2f_{\text{ISCO}}}{(1+z)}\right)^{-7/3} S_n\left(\frac{2f_{\text{ISCO}}}{(1+z)}\right)^{-1}\frac{2f_{\text{ISCO}}}{(1+z)}\,.
\end{align}
Here $p(m_{1}, m_{2})$ is the probability density function (PDF) of the source-frame masses and $\mathcal{A}$ depends on $r$ through the redshift $z=z(r)$. We have checked that $\mathcal{A}$ is a quantity of order $\mathcal{O}(1)$.

We observe that in Eq.~\eqref{eq:dipole}, the term proportional to $s$ is relevant only if a significant part of the population is across threshold. 
This term is due to the third line of Eq.~\eqref{eq:Npert}.
For a population of sources where $s\simeq 0$, i.e.\ such that nearly all sources are detected, then $\alpha(r) \simeq 2$, i.e. it is a fixed constant, independent on distance. 

In this case, the dipole can be used as a direct estimator of the observer velocity. More precisely, for a perfectly isotropic distribution of sources, i.e. with PDF $P(\Omega)=1/4\pi$, we can build the following observable (choosing $v_0$ aligned along the azimutal axis)
\begin{equation}
v_{{\bf n}'}\equiv \frac32\int {\rm d}\Omega\ P(\Omega)\ \left({\bf n} \cdot {\bf n}'\right)\ \mathcal{D}\left[{\bf n}\cdot {\bf v}_0\right]=\frac{\alpha v_0}{2c} \cos\theta'\simeq \frac{v_0}{c} \cos\theta'\,,
\label{eq:observable}
\end{equation}
where ${\bf n}'\equiv\left(\sin\theta'\ \cos\phi'\,,\sin\theta'\ \sin\phi'\,,\cos\theta'\right)$ is a vector pointing towards a generic fixed direction.
This observable is maximized when evaluated along the a priori unknown dipole direction and is exactly equal to the observer velocity, $v_0/c$, at the maximum.

If the term $s$ is not negligible, it is necessary to estimate it accurately in order not to bias the measurement of the observer velocity. One possibility is to measure $s$ directly from the catalog of events. Since we work at linear order in the observer velocity, $s$ depends only on the isotropic distribution of sources. Using Eq.~\eqref{eq:s} we can therefore write at zeroth order in $v_0/c$
\begin{align}
s(\rho_*,r)\equiv -\frac{\partial\ln \left(\frac{d N_{\rm det}(r,\rho>\rho_*)}{d\Omega dr}\right)}{2\partial\ln \rho_*}\, .  
\end{align}
To measure $s$ one can bin the events in bins of comoving distance (using a fiducial cosmology to translate the measured luminosity distance into $r$), and in bins of SNR, $\rho$. The quantity $s$ is then given by the slope of the cumulative number of events above $\rho$, evaluated at the chosen value $\rho_*$. This method has been used for example in~\cite{Euclid:2021rez} for galaxy number counts, providing a measurement of $s$ from the Euclid flagship simulation. Similarly the quantity $\mathcal{A}(r)$ can be measured from the catalog of events, using a fiducial cosmology to compute the probability distribution in the source frame mass from the distribution of masses in the observer frame. This may however be non-trivial and it requires also to account for non-trivial detector sensitivities as a function of frequency.

A complementary possibility would be to study the effect of $s$ with Monte Carlo simulations. For instance, as we show in Sec.~\ref{sec:4}, even if we are not able to detect all the BBHs simulated, the effect of $s$ does not seem to introduce a bias in the estimation of ${\bf v}_0$. Using Monte Carlo simulations, we could make sure that the effect of $s$ on $\alpha$ is lower than the typical statistical uncertainties on the estimation of the dipole which should scale as $1/\sqrt{N_{\rm tot}}$. Let us also recall that $s$ would impact the estimation of the kinematic dipole, but not its detectability with a given number of detections.

\subsection{A statistical estimator for the cosmic dipole}
\label{sec:2.3}

Let us now build an estimator for the observable $v_{{\bf n}'}$. We divide the sky in $N_{\rm sky}$ pixels of same solid angle, and associate a vector $\bn_i$ pointing to the center of each pixel. The estimator is given by
\begin{equation}
\label{eq:estimator}
\hat{v}_{{\bf n}'}=\frac3{2 N_{\rm tot}}\sum_{i=1}^{N_{\rm sky}} N_{\rm det}^{i}\cdot({\bf n}_i\cdot {\bf n}')\,,
\end{equation}
where $N_{\rm tot}$ is the total number of events and $N_{\rm det}^{i}$ is the number of sources that falls in the sky pixel $i$. This number can be written as
\begin{align}
\label{eq:Ni}
N_{\rm det}^{i}=\bar{N}_{\rm det}\left(1+\alpha\,\bn_i\cdot\frac{\bv_o}{c} \right) +\Delta N^i\,,
\end{align}
where $\bar{N}_{\rm det}=N_{\rm tot}/N_{\rm sky}$ is the mean number of events per pixel, and
$\Delta N^i$ accounts for the fact that 
the actual distribution of detected GW signals will not be exactly isotropic (even in the absence of a dipole), because of the stochastic nature of the coalescence distribution and of the detection process. Inserting~\eqref{eq:Ni} in~\eqref{eq:estimator}, the expected value of the estimator becomes
\begin{align}
 \langle\hat{v}_{{\bf n}'}\rangle&= \frac3{2 N_{\rm sky}}\alpha \frac{v_o}{c}\sum_{i=1}^{N_{\rm sky}}({\bf n}_i\cdot \hat{\bv}_o)({\bf n}_i\cdot {\bf n}')=\frac{\alpha v_o}{2c} \cos(\theta')=v_{\bf n'}\, ,\label{eq:exp}
\end{align}
because the isotropic part of the actual sky distribution averages to zero due to the $({\bf n}_i\cdot {\bf n}')$ factor, and the expectation value of the stochastic noise is, by definition, zero. Hence  what survives is exactly the dipole component discussed in the previous section, which reduces to the observer velocity when $\alpha=2$, i.e.\ when threshold effects are negligible.

In order to assess the statistical detectability of the dipole, one must determine the variance of the estimator, due to the stochastic fluctuations of events, $\Delta N^i$, around the monopole.
Since the pixels have associated the same solid angle, the number of detections contained in every pixel is drawn from the same distribution, that is a Poissonian with $\mu=\bar{N}_{\rm det}$. One can then determine the variance of $\hat{v}_{{\bf n}'}$ just by summing in quadrature the variances within each pixel: 

\begin{equation}
\label{eq:variance}
    \langle\Delta \hat{v}_{{\bf n}'}^2\rangle=\frac{9}{4N_{\rm tot}^2}\sum_{i=1}^{N_{\rm sky}}\bar{N}_{\rm det}\cos^2{\theta'}_i
    \simeq\frac{3}{4N_{\rm tot}}\, ,
\end{equation}
as the sum in $\cos^2\theta_i$ quickly converges to $N_{\rm sky}/3$, already for $N_{\rm sky}< 10$. We thus have the intuitive result that the precision attainable in the dipole measurement scales down with the square root of the total number of detections, as expected in an essentially poissonian process.

From  Eq.~\eqref{eq:variance}, one expects that in order to detect a velocity dipole $v_0/c$ of order $10^{-3}$,
one needs a very large number of GW detections, $N_{\rm tot}\simeq 10^6$, which can be reached only with XG detectors.

Another way to obtain the same result is to artificially generate a large number of random sky distributions, for instance by randomly reshuffling the positions of the detected GW events, and compute the variance of the estimator on such a collection of sky distributions. As will be shown in the next section, the result of this estimate is in very good agreement with the simple poissonian computation outlined above.

\section{Forecasts with XG GW detectors}
\label{sec:3}
In this section, we forecast the detectability of the cosmic dipole with GW events. In Sec.~\ref{sec:sim} we start by discussing the detection prospects for compact binary mergers with XG detectors. In Sec.~\ref{sec:cdet} we study the significance of a possible cosmic dipole detection using the estimator in Eq.~\eqref{eq:estimator}. Finally, in Sec.~\ref{sec:cbayes} we show an example of  how to estimate the cosmic dipole using Bayesian statistic.

\subsection{Simulating compact binaries mergers with XG detectors}
\label{sec:sim}

We consider a network of XG detectors composed by ET \citep{2010CQGra..27s4002P} and two CE \citep{PhysRevD.91.082001,Reitze:2019iox}. For ET we assume the same PSD used in \citet{2022arXiv220702771I}, while for CE we use the PSD provided by the CE consortium\footnote{\url{https://cosmicexplorer.org/sensitivity.html}}. In this exploratory approach, we detect a GW event using a SNR threshold calculated with Eq.~\eqref{eq:SNR1}\footnote{This is a reasonable approximation for BNS events, but not so much for BBHs; however in the latter case threshold effects are less important (see end of this section) and this crude approximation should not affect our result.}. As we detect the time-evolution of the antenna patterns of the GW detectors, we set a lower frequency cut-off in Eq.~\eqref{f731} at $5$ Hz.
Given the sensitivity curves of the network and the two values of the source-frame masses, Eq.~\eqref{eq:SNR1} can be used to calculate the maximum redshift at which we will be able to observe binaries with a given SNR as threshold.
\begin{figure}
    \centering
    \includegraphics{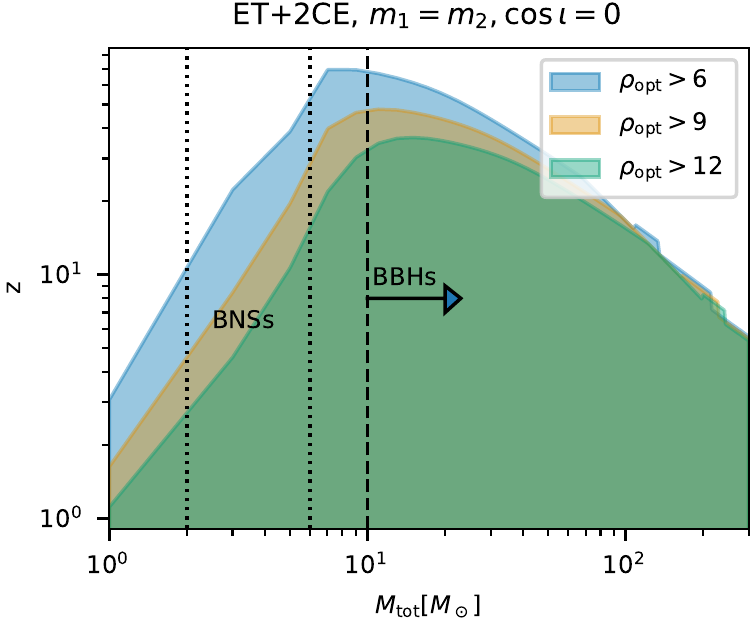}
    \caption{Detection horizons for a ET+2CE detectors network. The redshift horizons (vertical axis) are calculated as a function of total source-frame mass (horizontal axis) for an equal mass binary ``edge-on'' with respect to the observer (worst-case scenario). The different colors mark the horizons for 3 SNR thresholds of 6, 9 and 12. Detection horizons are calculated using a flat $\Lambda$CDM cosmology with $H_0=67.7 \mathrm{km \, s^{-1} \, Mpc^{-1}}$ and $\Omega_m=0.3097$. Note that we only consider binaries with \textsc{ISCO} frequency above $5$~Hz for our simulation.}
    \label{fig:horizons}
\end{figure}

Fig.~\ref{fig:horizons} reports the maximum redshift, as a function of total source frame mass, up to which we will be able to detect compact binary coalescences with a SNR $>6,9$ and $12$. We note that the horizons reported in Fig.~\ref{fig:horizons} are calculated for edge-on binaries ($\cos \iota=0$) and therefore they represent the worst-case scenario at which we will be able to detect compact binaries. Fig.~\ref{fig:horizons} indicates that if we consider a SNR for detection of $9$, we will be able to detect all the BNSs merging below redshift $\sim 0.8$ and all the BBHs merging below redshift $\sim 2$. 

BBHs are more massive than BNSs, which is why we expect them to have a louder SNR and a higher detection range (see Eq.~\eqref{eq:SNR1}). However, when a BBH is too massive, it merges at low frequencies (see Eq.~\eqref{eq:isco}) thus spending a small number of cycles in the sensitivity band of GW detectors and collecting less SNR. Moreover, low-frequency regions are not very sensitive for ground-based GW detectors \citep{2020JCAP...03..050M,2022arXiv220702771I}. That is why in Fig.~\ref{fig:horizons}, the detection range starts to decrease after $\sim 40 M_\odot$. This behavior mostly depends on the GW detector's sensitivity as a function of frequency. Ground-based GW detectors usually have the same design for sensitivity as a function of frequency. While we might change BBHs population models, detector sensitivities and other simulation prescriptions, recent works indicate that we expect to detect almost all the population of BBHs merging in the Universe with next-generation detectors \citep{2020JCAP...03..050M,2022arXiv220702771I}.

We want to determine if these detection horizons can introduce a significant selection bias when detecting populations of BNSs and BBHs. If a strong selection bias is present, then the theoretical prediction for the dipole depends on the parameter $s(\rho_*,r)$ in Eq.~\eqref{eq:defs}, that we need to model. If no strong selection bias is induced, i.e.\ if $s(\rho_*,r)\simeq 0$, then $\alpha\simeq 2$ and our estimator provides a model-independent measurement of the observer velocity $v_o$. 
\begin{figure}
    \centering
    \includegraphics{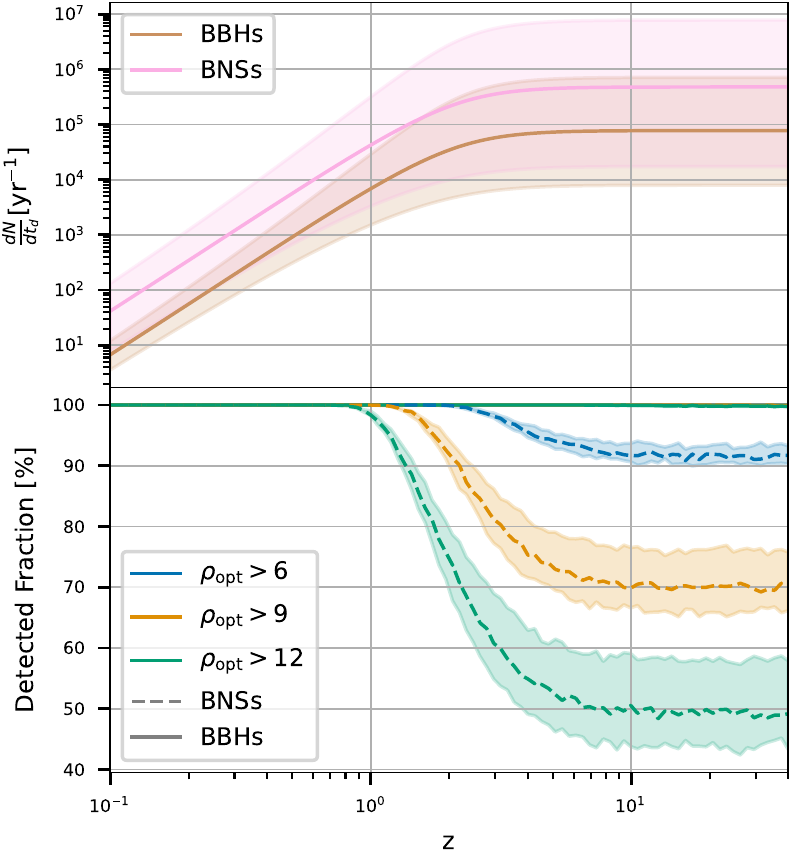}
    \caption{\textit{Top panel}: Total number (vertical axis) of GWs from BNS and BBH mergers arriving on Earth within a given redshift shell (horizontal axis). The shaded areas mark the 90\% credible intervals associated to the population model, while the solid lines their median values. \textit{Bottom panel}: Total fraction of GWs that will be detectable (with various SNR thresholds) within a given redshift shell.
    The figures are generated using the same population model for BNS and BBHs as in \citet{2022arXiv220702771I}. The solid lines indicate the fiducial model for the merger rates, while the shaded area the contours identified by the 90\% credible interval uncertainties on the rate models.}
    \label{fig:rates}
\end{figure}

We simulate BNSs and BBHs following the same population models for BNSs and BBHs reported in \citet{2022arXiv220702771I} and supported by current observations \citep{2021ApJ...913L...7A}. More details about the BNSs and BBHs mass distributions are given in App.~\ref{app:pop}. We use the same merger rate model as a function of redshift for both BNSs and BBHs, given by, see~\citet{2020ApJ...896L..32C}
\begin{equation}
    R(z)=R_0 [1+(1+z_p)^{-\gamma-k}] \frac{(1+z)^\gamma}{1+\left(\frac{1+z}{1+z_p}\right)^{\gamma+k}}\,,
    \label{eq:rate}
\end{equation}
where $R_0$ is the merger rate today, and $\gamma$ and $k$ are two parameters encoding the redshift evolution of the merger rate. Since $k$ and $\gamma$ are not constrained by current observations, we assume fiducial values of $k=3$ and $z_p=2$, which are consistent with the Star Formation Rate \citep{2014ARA&A..52..415M}. For $\gamma$ and $R_0$ we take values consistent with the 90\% credible intervals values found by \citet{2021arXiv211103634T}.  For BBHs we take $R_{0,\rm BBH}=17^{+10}_{-6.7} {\rm Gpc^{-3} yr^{-1} }$ and  $\gamma= 2.7^{+1.7}_{-1.8}$, while for BNSs we take $R_{0,\rm BNS}=106^{+190}_{-93} {\rm Gpc^{-3} yr^{-1} }$ and the same $\gamma$ as for BBHs (since it was not possible to constrain this parameter from current observations). We take as fiducial rate models the ones corresponding to the median values of the parameters.
The number of GWs, emitted within a redshift shell with $z'<z$, crossing the Earth per year is then given by
\begin{equation}
    \frac{dN}{dt_d}= \int_0^{z} R(z')\frac{1}{1+z'} \frac{dV_c}{dz'} dz'\,,
\end{equation}
where $dV_c/dz$ is the differential of the comoving volume and $t_d$ is the detector time in years.
In the top panel of Fig.~\ref{fig:rates} we report the total number of GWs from BNS and BBH mergers arriving in 1 year of observing time. With the assumed model for the merger rate of BNSs and BBHs we find that the total number of BNS and BBH mergers saturates at redshift $\sim 2$. This is expected from the assumed merger rate model that has a peak around redshift $2$. If we consider all the observable Universe, we find that we might expect to have between $(1-50) \cdot 10^4$ GWs from BBHs arriving on Earth per year and $10^4-10^7$ GWs from BNSs per year. However, given the detection horizons in Fig.~\ref{fig:horizons} not all the GWs will be detectable. 

The bottom panel of Fig.~\ref{fig:rates} shows the ``\textit{detectable fraction}'' of BNS and BBH mergers within a certain redshift. On one hand, the plot shows that we will be basically able to detect all the BBH mergers in the Universe. This means that all events are above threshold, leading to $s(\rho_*,r)\simeq 0$ at all distances. For BNSs, on the other hand, we see that above $z\simeq 1$, a sizeable fraction of events will not be detectable. As a consequence $s(\rho_*,r)$ may be large, leading to a non-negligible contribution of threshold effects to the parameter $\alpha$. Therefore, even though we expect significantly less detections from BBHs than from BNSs, the former will offer a cleaner measure of the observer velocity, which is independent of the characteristics of the population.

Using the merger rate prescription in Eq.~\eqref{eq:rate} we also estimate the clustering dipole contribution following App.~A of \citet {2021ApJ...908L..51S}. To calculate the clustering dipole contribution, we use \textsc{nbodykit} \citep{Hand:2017pqn} to obtain the matter density power spectrum at $z=0$ using the Halofit \citep{Takahashi:2012em} prescription to include non-linearities in the CLASS code \citep{2011arXiv1104.2932L} for a Planck15 cosmology. By using the power spectrum at $z=0$ we are overestimating the clustering dipole (since the power spectrum decreases with redshift) and we therefore obtain a conservative estimation for its importance with respect to the kinematic dipole.
We obtain a value of the clustering dipole of $\mathcal{D}_{\rm cl} \approx 1.2 \cdot 10^{-4}$ which is about one order of magnitude lower than the simulated cosmic kinematic dipole for AGN and CMB.

\begin{figure}
    \centering
    \includegraphics[scale=0.35]{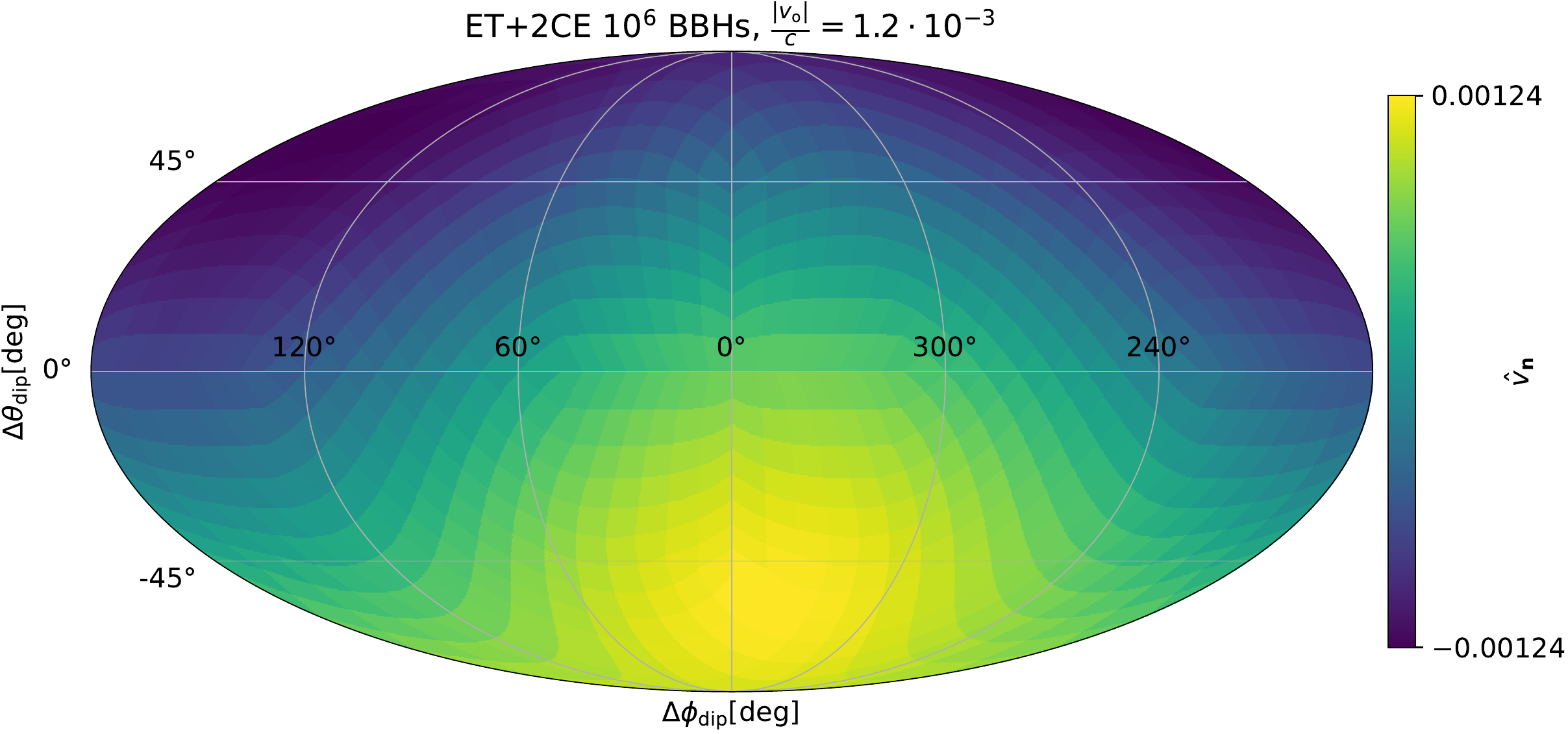}
    \includegraphics[scale=0.35]{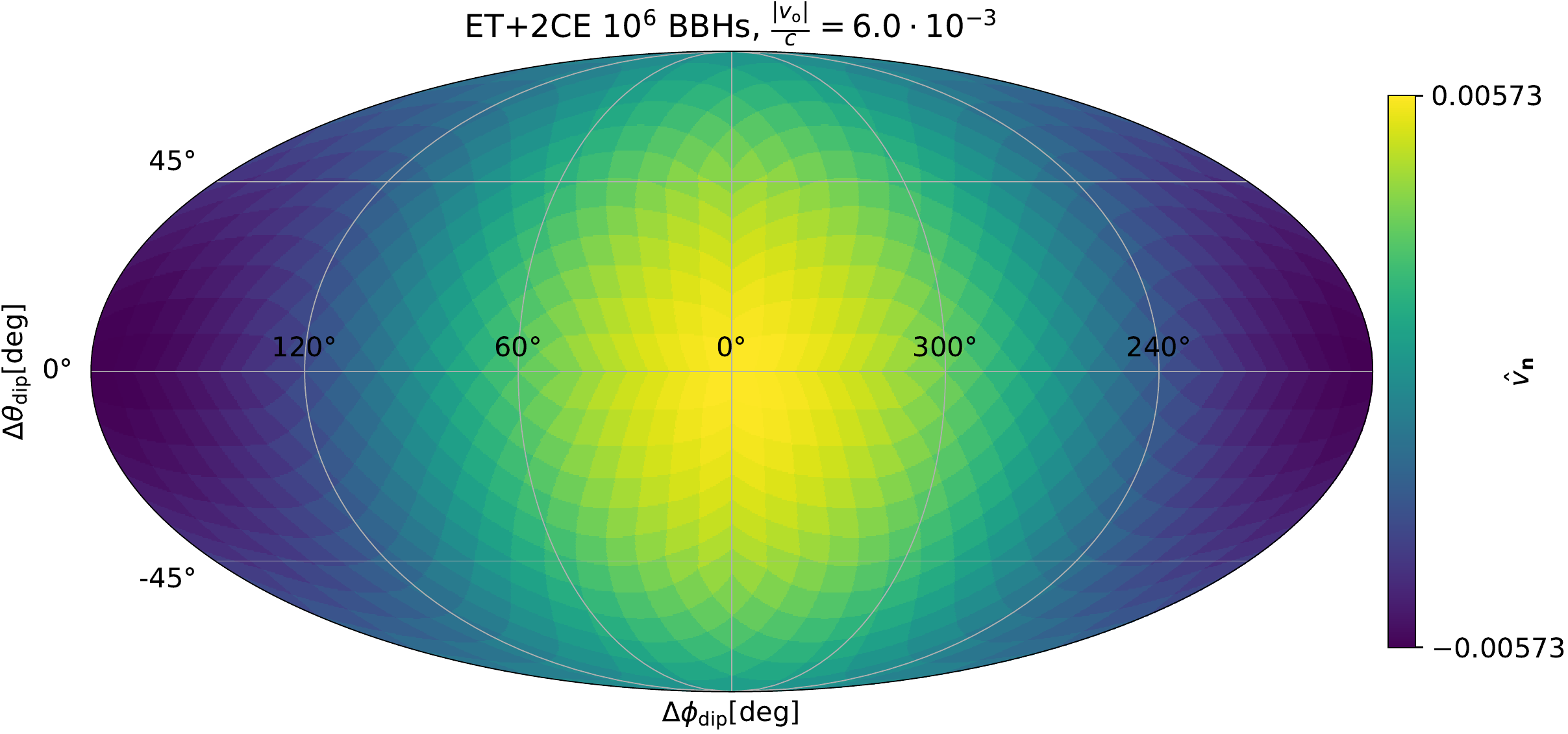}
    \caption{Sky distribution of the estimator $\hat{v}_{\mathbf{n}}$ for the CMB fiducial value  (top map) and the AGN fiducial value (bottom map). The maps are centered around the injected direction of the observer velocity. The figure is generated by dividing the sky in equal size pixels of 53 deg$^2$.}
    \label{fig:maps}
\end{figure}

\subsection{Detection and Estimation of the cosmic dipole}
\label{sec:cdet}

In this section we study the detectability of the GW dipole, considering two fiducial values for its amplitude. The first one assumes that the GW dipole is of purely kinematic nature, with a velocity consistent with the one inferred from CMB observations \citep{2020A&A...641A...6P}, i.e.\ $v^{\rm CMB}_{\hat{\bv}_o}=v_{o,{\rm CMB}}/c=1.2 \cdot 10^{-3}$. Since observations from radio sources and quasars find a dipole which is 2-5 times larger than expected~\citep{2017MNRAS.471.1045C,2018JCAP...04..031B,2021ApJ...908L..51S,2021A&A...653A...9S,Secrest:2022uvx}, we also consider a second fiducial value for the GW dipole, that would be consistent with these observations. Indeed, if the large dipole found in these studies is due to a breaking of the Copernician principle~\citep{Secrest:2022uvx}, i.e.\ to an instrinsic large anisotropy in the distribution of structures, then this anisotropy should be present also in the distribution of GW sources, that follow the large-scale structure of the Universe. It is thus interesting to assess the detectability of such a large dipole with GW sources. For this we take the extreme case of a dipole which would be 5 times larger than the one expected from CMB velocity, and with a direction aligned with it, i.e. $v^{\rm AGN}_{\hat{\bv}_o}=6.0 \cdot 10^{-3}$.

For each fiducial value of the dipole, we simulate $10^4,10^5,10^6$ and $10^7$ BBHs detections $N_{\rm tot}$, using the ET+2CE network with a SNR thresold of $9$ for detection and the population of BBHs described in the previous section. The BBH detections are simulated as follow. 
Each simulated BBH mass and redshift are drawn from the distributions described in the previous section and in App.~\ref{app:pop}. The original distribution of BBHs is isotropic in sky. Then, for each BBHs, we introduce the effect of the observer velocity by introducing an aberration following Eq.~\eqref{eq:domegabar}: $\theta'=\theta-(v_o/c)\sin \theta$, where $\theta$ is the angle between the source position and the observer velocity~\footnote{Note that in the AGN case, the dipole may not be due solely to the observer velocity, but rather to a large anisotropy in the distribution of sources, as discussed above. However, when assessing the detectability of such a dipole, it does not matter if it is of kinematic origin or not. Hence, we can simply simulate it as if it were due to a large velocity, 5 times larger than the one measured from the CMB.}. The detector-frame mass and luminosity distance change as described in Eq.~\eqref{eq:pert}. For each BBH, we then calculate the SNR using Eq.~\eqref{eq:SNR1}. Then an ``observed'' SNR is drawn from a $\chi^2$ distribution with 2 times number of detectors d.o.f. If the ``observed'' SNR exceeds a SNR threshold of $9$, the binary is labelled as detected. Finally, to mimic the sky localization uncertainty given by the GW detection, we scatter the position of the GW sources using a gaussian distribution with $3$ deg. 

Once the total list of BBHs detections is obtained with their sky position, we calculate the estimator of the GW dipole defined in Eq.~\eqref{eq:estimator}. Fig.~\ref{fig:maps} shows the skymap of $\hat{v}_{\mathbf{n}}$ calculated for the CMB and AGN  fiducial values using $10^6$ BBHs detections. As we can see from the figure, in the AGN case, $\hat{v}_{\mathbf{n}}$  is maximized in the direction of the observer velocity and it displays an amplitude of $\hat{v}_{\mathbf{n}}=5.73 \cdot 10^{-3}$, similar to the one injected. For the CMB case on the other hand, the estimator is maximized in a direction different from the one of the observer velocity and it displays a maximum value of $\hat{v}_{\mathbf{n}}=1.3 \cdot 10^{-3}$, higher than the one injected. The reason for this is that, in the CMB case, the GW  dipole will only be marginally detectable  with $10^6$ BBHs.
\begin{figure}
    \centering
    \includegraphics{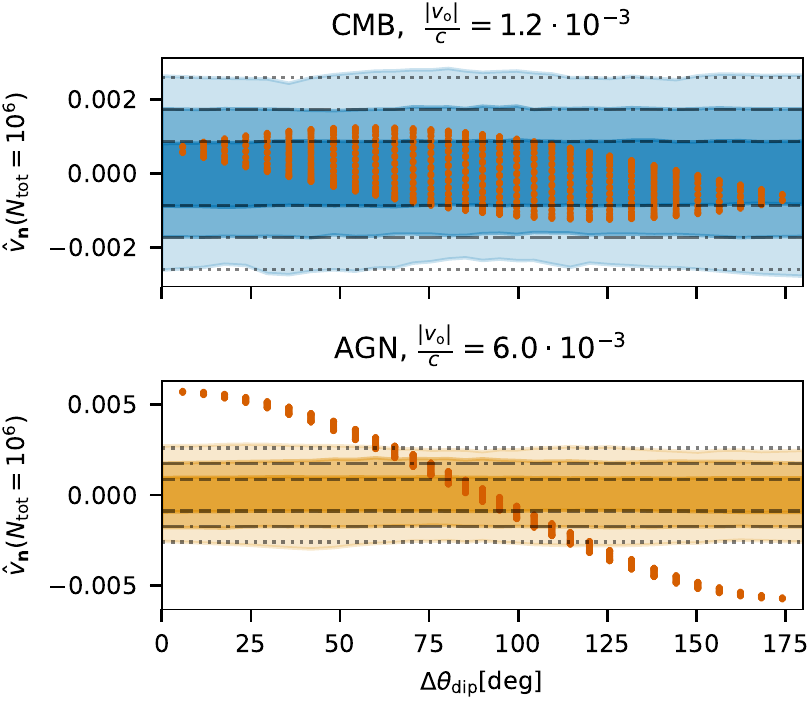}
    \caption{Distribution of the estimator $\hat{v}_{\mathbf{n}}$ (red points) for the CMB fiducial value (top plot) and the AGN fiducial value (bottom plot), plotted as a function of the angle between $\bn$ and the injected velocity direction $\hat{\bv}_o$. The shaded areas mark the background $1,2,3 \sigma$ confidence intervals generated by randomly shuffling the detections over the sky. The horizontal dashed lines mark the standard deviations generated with Eq.~(\ref{eq:variance}). 
    The figure is generated by dividing the sky equal size pixels of 53 deg$^2$.}
    \label{fig:estimator_vs_angle}
\end{figure}

Fig.~\ref{fig:estimator_vs_angle} shows another view of the values of the estimator reported in Fig.~\ref{fig:maps}. More precisely, we plot the value of the estimator as a function of the angle between the direction of the observer velocity, $\hat{\bv}_o$, and the direction $\mathbf{n}$ at which the estimator is calculated. The figure also displays the $1, 2$ and $3$ $\sigma$ values of $\hat{v}_{\mathbf{n}}$ due to the stochastic fluctuations of events around the monopole  given by Eq.~\eqref{eq:variance}. A $3\sigma$ detection of the GW dipole can be claimed if the $\hat{v}_{\mathbf{n}}$ estimator exceeds the $3 \sigma$ threshold from Poisson noise. From the figure we can see that, with $10^6$ detections, a GW dipole with amplitude compatible with that of the CMB dipole is marginally detectable with $1\sigma$ confidence (the probability that it is generated from a fluctuation of the monopole is not negligible), while a GW dipole with amplitude compatible with that of the AGN dipole is clearly detectable.
\begin{figure}
    \centering
    \includegraphics{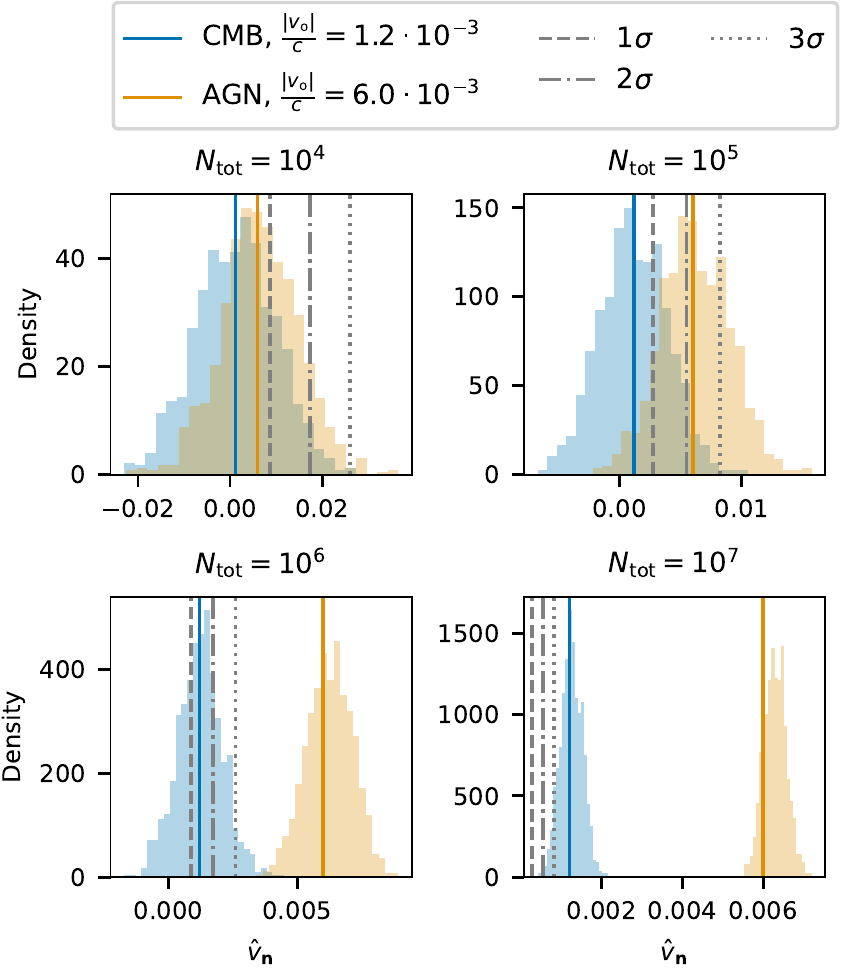}
    \caption{Distribution of the estimator $\hat{v}_{\mathbf{n}}$ evaluated in the injected velocity direction $\bn=\hat{\bv}_o$ with $10^4,10^5,10^6$ and $10^7$ BBH detections. The histograms are obtained by simulating 1000 populations of BBHs. The vertical dashed lines indicate the fiducial values of the GW dipole in the CMB case (blue) and the AGN case (orange). The dashed and dotted gray lines indicate the $1,2,3 \sigma$ contribution from Poisson noise, generated using Eq.~(\ref{eq:variance}).}
    \label{fig:detectability}
\end{figure}

In order to better understand the detectability of the GW dipole, we repeat 1000 times the previous simulations for $10^4,10^5,10^6$ and $10^7$ detected BBHs and we calculate the fraction of cases for which the dipole will be detected. Fig.~\ref{fig:detectability} shows the distribution of the estimator  $\hat{v}_{\mathbf{n}}$, evaluated in the velocity direction $\bn=\hat{\bv}_o$, that we obtain for the CMB and AGN cases. From the figure we can see that the distributions are centered around the injected value of the dipole. This is a confirmation that, for BBHs, threshold effects are negligible and $\alpha=2$ (such that $\alpha/2=1$ in Eq.~\eqref{eq:exp}). From the figure, we can see that when we only have $10^4$ detections, the distribution of the dipole estimator for the AGN and the CMB cases almost coincide. Moreover, the $1,2,3 \sigma$ intervals of these distributions almost coincide with the $1,2,3 \sigma$ variance from Poisson noise. In other words, with only $10^4$ detections the dipole is not detectable. As we collect more and more detections, the dipole estimator distributions tend to become separate from the Poisson noise.
\begin{figure}
    \centering
    \includegraphics{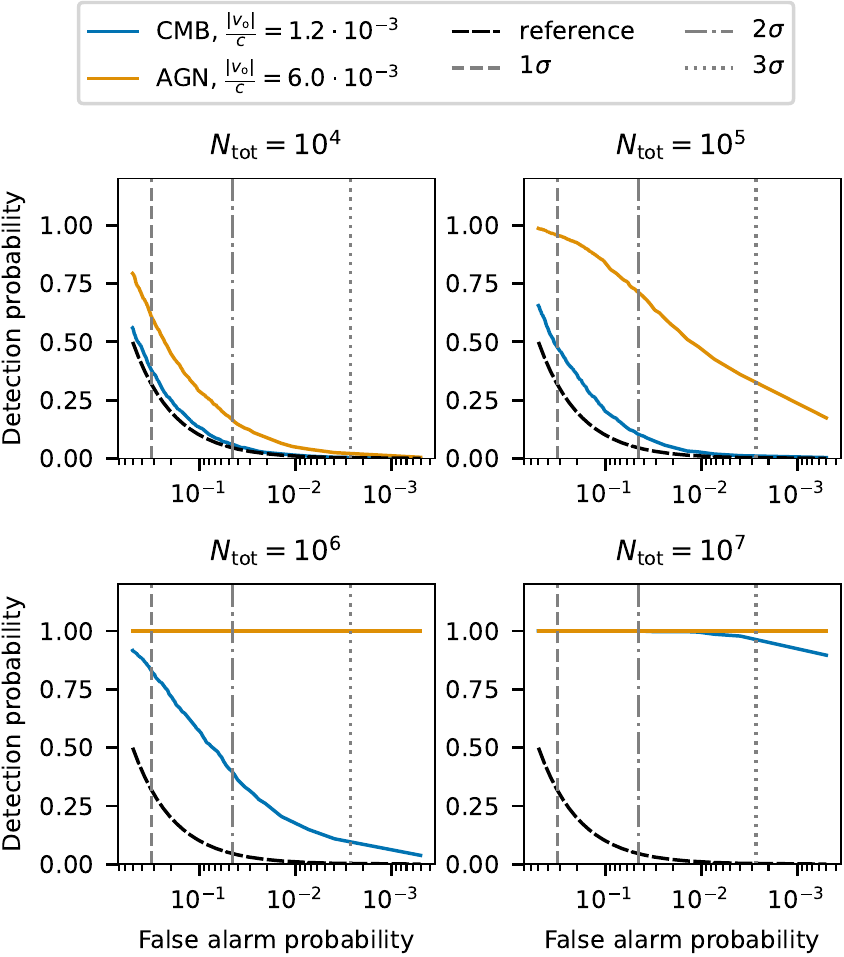}
    \caption{FAP versus detection probability plots for the injected CMB and AGN fiducial values (blue and orange lines). The vertical dashed lines indicate the FAP at the standard $1,2,3 \sigma$ credible intervals. The black dashed line is the reference to indicate when the detection probability coincides with the FAP (they are the same distribution).}
    \label{fig:fappdet}
\end{figure}

In order to better quantify the detectability of the dipole, we plot False Alarm probability (FAP) vs detection probability in Fig.~\ref{fig:fappdet}. The FAP identifies a threshold for the dipole detection and it is defined as the probability that a random fluctuation of the number of GW detections in absence of dipole, would result in a false positive. The detection probability is defined as the probability that, in presence of a dipole, the estimator for the dipole detection would exceed the FAP threshold. Decreasing the FAP allows us to be more sure on the dipole nature of our detection but it decreases our sensitivity for the dipole detection. FAP versus detection probabilities plots can be used to: \textit{(i)} estimate what are the detection prospects for a given threshold (significance of the detection) and \textit{(ii)} clearly check in what regime the dipole is detectable. In fact, in the case where we are not able to detect the dipole, we expect the detection probability to be equal to the FAP, as the detection estimator would follow the same distribution.

As we can see from the top left panel of  Fig.~\ref{fig:detectability}, with $10^4$ GWs detections, the distribution of the estimator for the CMB case agrees with the confidence intervals traced by the detection thresholds. This is reflected in the top left panel of Fig.~\ref{fig:fappdet}, where we see that the FAP and detection probability follow the same distribution, thus indicating that the only dipole that can be detected in this case, is the one from random fluctuations around the isotropic background (false positive). For the AGN case, the detection probability is slightly larger than the FAP, but still no robust detection could be claimed in this case. 
With $10^5$ GWs detections ($\sim$ 2-3 year of observation), we can see from Fig.~~\ref{fig:fappdet}  that there is a 75\% probability of detecting a dipole with AGN amplitude using a FAP of $2\sigma$. With $10^6$ detections (achievable in 10 years for an optimistic scenario), a dipole as high as the AGN one would be detectable at 100\%. On the other hand, if the GW dipole is compatible with the one observed in the CMB, then there is a 50\% probability that we would detect it with a FAP of $2 \sigma$.

\subsection{A Bayesian study}
\label{sec:cbayes}
Bayesian statistic can also be used to provide evidence for the GW dipole and estimate its parameters. As discussed in Section~\ref{sec:2.1}, the number of observed BBHs in a pixel situated in direction $\bn_i$  is given by  
\begin{align}
N_{\rm det}^{i}=\bar{N}_{\rm det}\left(1+\alpha\,\bn_i\cdot \frac{\bv_o}{c} \right) +\Delta N^i\, , \label{eq:Ni2}
\end{align}
where $\bar{N}_{\rm det}$ is the number of detections per pixel due to the monopole, $\Delta N^i$ a statistical fluctuation ,  and $\alpha \approx 2$ since for BBHs threshold effects are negligible. 
The likelihood of obtaining $k_i$ detections in one pixel $\bn_i$ is then given by
\begin{equation}
    \mathcal{L}\left(k_i|\bar{N}_{\rm det},\frac{\bv_o \alpha}{c}, \bn_i \right) \propto e^{-N^i_{\rm det}}\cdot \left(N^i_{\rm det}\right)^{k_i}\, .
\end{equation}
The overall likelihood of obtaining $\{k\}=\{k_1,\ldots,k_{N_{\rm sky}}\}$ detections when dividing the sky in $N_{\rm sky}$ equal area pixels is
\begin{equation}
    \mathcal{L}\left(\{k\}|\bar{N}_{\rm det},\frac{\bv_o \alpha}{c}\right) = \prod_{i}^{N_{\rm sky}} \mathcal{L}\left(k_i|\bar{N}_{\rm det},\frac{\bv_o \alpha}{c}, \bn_i\right)\,.
    \label{eq:likelihoodtotal}
\end{equation}

Finally, by applying the Bayes theorem, we can obtain posterior distributions on $\bar{N}_{\rm det},\alpha v_o/c$ and $\hat{\bv}_o$ by calculating
\begin{align}
    p\left(\bar{N}_{\rm det},\frac{\bv_o \alpha}{c}| \{k\} \right) &\propto  \mathcal{L}\left(\{k\}|\bar{N}_{\rm det},\frac{\bv_o \alpha}{c}, \right)\times \pi \left(\bar{N}_{\rm det},\frac{\bv_o \alpha}{c}, \right)\,,
\end{align}
where $\pi(\cdot)$ is a prior term.

We implement the likelihood in Eq.~\eqref{eq:likelihoodtotal} in a nested sampling code to obtain posterior distributions for $N_{\rm tot}=\bar{N}_{\rm det} N_{\rm sky}$ and $\alpha v_o/c$, in the case studies of the CMB and AGN dipole amplitudes, estimated with $10^6$ GWs. We use the python code \textsc{bilby} \citep{2019ApJS..241...27A,2020MNRAS.499.3295R} and its implementations of the nested sampling algorithm  \textsc{dynesty} \citep{2019S&C....29..891H}. We use an isotropic prior for the dipole direction, a flat in log prior for $\alpha v_o/c$ between $10^{-6}$ and $10^{-1}$ and a flat in log prior for the total number of events $N_{\rm tot}$ between $10^5$ and $10^8$.

\begin{figure}
    \centering
    \includegraphics[scale=0.35]{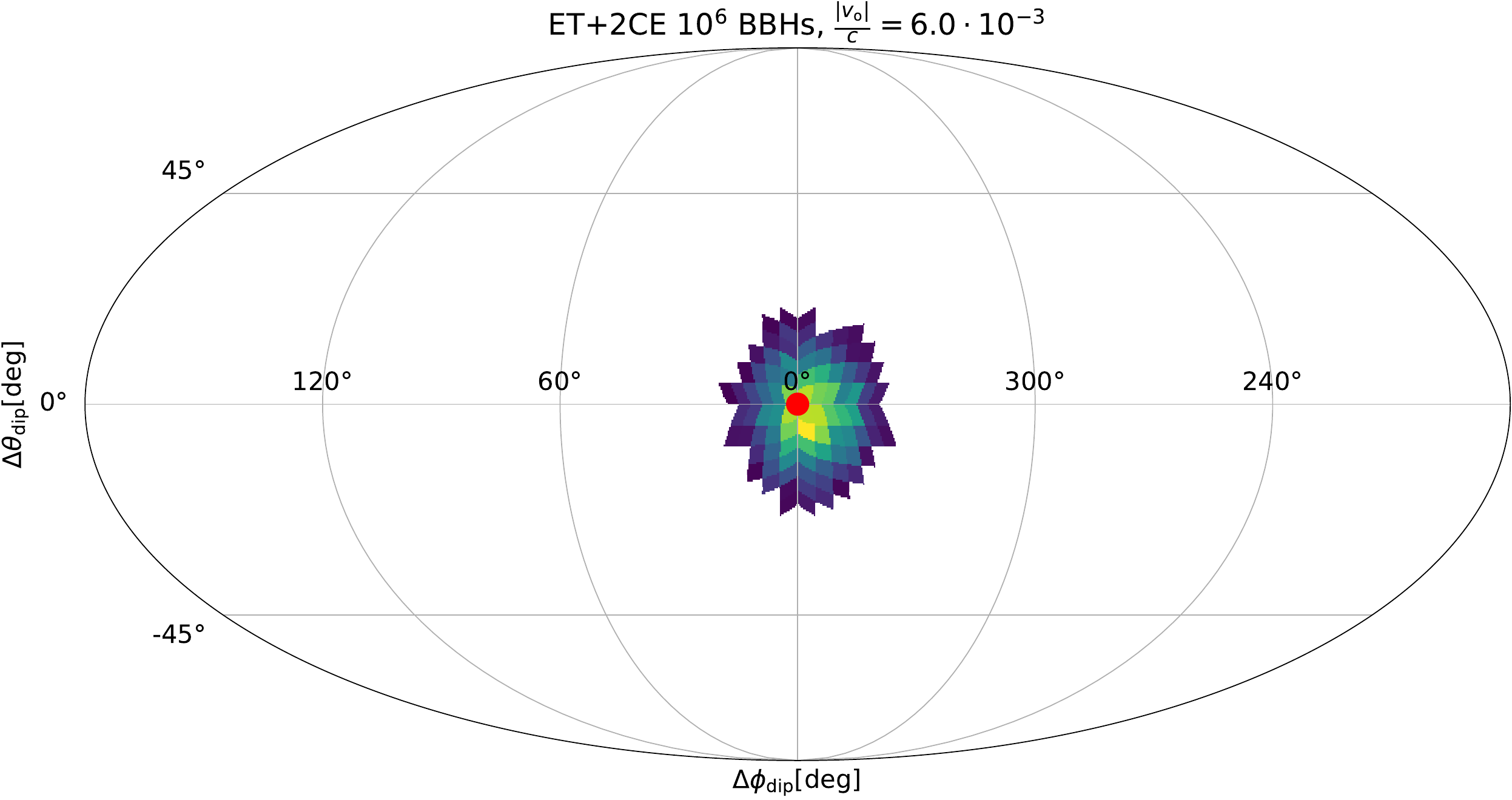}
    \caption{Sky area at 90\% credible intervals for the dipole sky direction for the AGN test case with $10^6$ detections. The red dot marks the direction of the dipole. The sky localization has a radius of $\sim$ 13 deg.}
    \label{fig:AGN_dipoleloc}
\end{figure}

For the case of a GW dipole with amplitude compatible with the AGN one, we find that with $10^6$ GWs detections, we are able to estimate the direction of the observer velocity with an uncertainty of $\sim 13$ deg ($\sim 2200$ deg$^2$ of area) at 90\% credible intervals. Fig.~\ref{fig:AGN_dipoleloc} shows the 90\% credible intervals in the sky identified by the posterior. 
\begin{figure}
    \centering
    \includegraphics[scale=0.5]{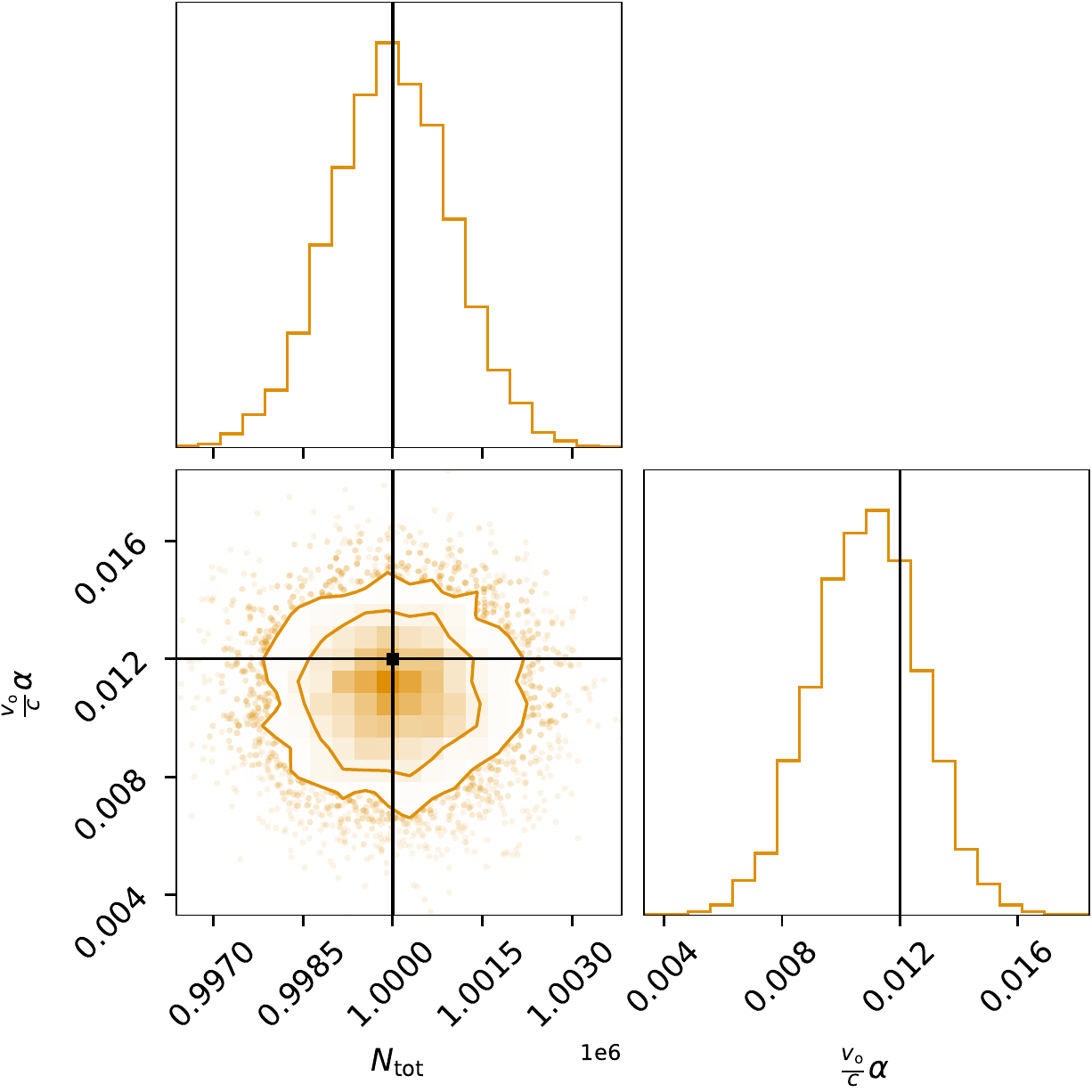}
    \caption{Plots of the posterior distribution (and its marginals) for the total number of detections $N_{\rm tot}$ and the dipole amplitude $\alpha v_o/c$ for the AGN test case with $10^6$ detections. The solid black lines mark the injected values (assuming $\alpha=2$). }
    \label{fig:amplitude_AGN}
\end{figure}
Fig.~\ref{fig:amplitude_AGN} shows instead the estimate of the total number of detections and the value of the GW dipole amplitude, $\alpha v_o/c$. The value of the recovered dipole amplitude is $\alpha v_o/c=1.15^{+0.18}_{-0.18} \cdot 10^{-2}$ at 68.3\% symmetric credible intervals. The amplitude of a dipole consistent with the AGN one can therefore be measured with a precision of 16\%. 

We can also calculate the Bayes factor to asses the detection of the dipole, i.e.
\begin{equation}
    \mathcal{B}^{\rm dip}_{\rm mono}= \frac{p(\rm dip|\{k\})}{p(\rm mono|\{k\})},
\end{equation}
where $p(\rm dip|\{k\})$ and $p(\rm mono|\{k\})$ are the evidences for the dipole model (with $\alpha v_o /c \neq 0$) and the monopole model (with $\alpha v_o/c =0$). For the AGN case, with $10^6$ GW detections we obtain a $\log_{10}(\mathcal{B}^{\rm dip}_{\rm mono})=6.0$, so strong preference for the presence of a dipole.

On the other hand, if the GW dipole has an amplitude consistent with that of the CMB dipole, the situation is different. First, we find that with $10^6$ detections it will not be possible to constrain the sky location. This is expected, since in Sec.~\ref{sec:cdet} we showed that the GW dipole would be hardly detectable in this case with $10^6$ events. 
\begin{figure}
    \centering
    \includegraphics[scale=0.5]{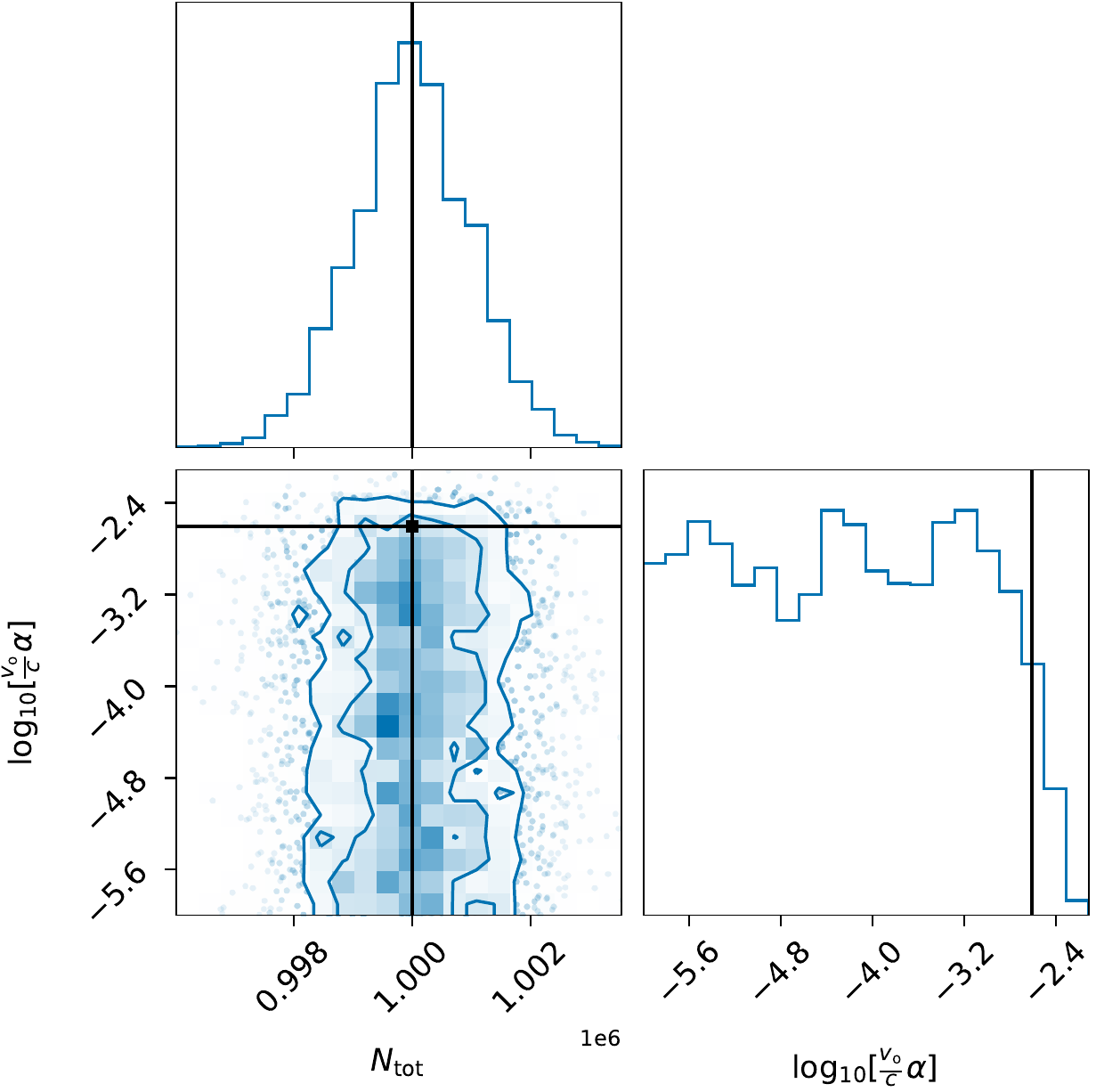}
    \caption{Plots of the posterior distribution (and its marginals) for the total number of detections $N_{\rm tot}$ and the dipole amplitude $\alpha v_o/c$ for the CMB test case with $10^6$ detections. The solid black lines mark the injected values (assuming $\alpha=2$).}
    \label{fig:amplitude_CMB}
\end{figure}
Moreover, if we look at the posteriors on the total number of events, $N_{\rm tot}$, and on the dipole amplitude, $\alpha v_o/c$, plotted in Fig.~\ref{fig:amplitude_CMB}, we can see that the total number of detections is clearly constrained, while the dipole amplitude cannot be constrained. However, it is possible to define an upper-limit in this case that results in $\alpha v_o/c <3.0 \cdot 10^{-3}$ at 95\% credible intervals (the injected value was $2.4 \cdot 10^{-3}$). We also compute the Bayes factor for this case finding that $\log_{10}(\mathcal{B}^{\rm  dip}_{\rm mono})=-0.17$, indicating that there is no clear preference for the presence of a dipole or not.

\section{Discussion}
\label{sec:4}
In the previous section we have discussed several aspects related to the detection and estimation of the cosmic dipole from GW counting using both frequentists and bayesian techniques. We have shown that BBHs detected with the ET+2CE might be a ``clean''  probe for estimation of the dipole, since threshold effects are negligible in this case and the dipole is consequently only due to aberration ($\alpha \approx 2$).

The first crucial task to consider when searching for a dipole with GW number counts is the assessment of the detection significance. 
For the frequentists techniques, we have shown that the significance can be evaluated using standard p-value techniques for Poissonian statistic and by reshuffling GW events in the sky to build a distribution of the noise due to the stochastic distribution of sources. For Bayesian statistic instead, we have shown that the detection can be evaluated with Bayes factors between the dipole and the monopole models. In both cases, we have obtained that in order to detect a dipole with amplitude $\alpha v_o/c$, one would need at least $\sim (c/\alpha v_o)^2$ detections, in order to be significantly confident that the detected dipole is not a fluctuation of the monopole distribution. We stress that, even with a few detections, it would not be surprising to find that the distribution of events is not perfectly isotropic in the sky. However, a low number of detections does not have the statistical significance to fit a dipole distribution. For instance, in Fig.~\ref{fig:maps}, we have shown that for a simulated GW dipole consistent with the one of the CMB, our estimator shows a peak over the sky which is not directly aligned with the simulated dipole. However in this case, its value is not statistically significant and a detection cannot be claimed. 

This could explain the puzzling results of \citet{2022arXiv220407472K}, where using LIGO/Virgo data the authors reconstruct a dipole direction orthogonal to the CMB one. In that work the authors indeed do find an asymmetry in the distribution of GW events over the sky, however they do not assign any statistical significance to their results. If a significance was worked out, it would very probably indicate that the dipole reconstruction is statistically not significant due to the very low statistics of events used in the analysis.
\begin{figure}
    \centering
    \includegraphics[scale=0.6]{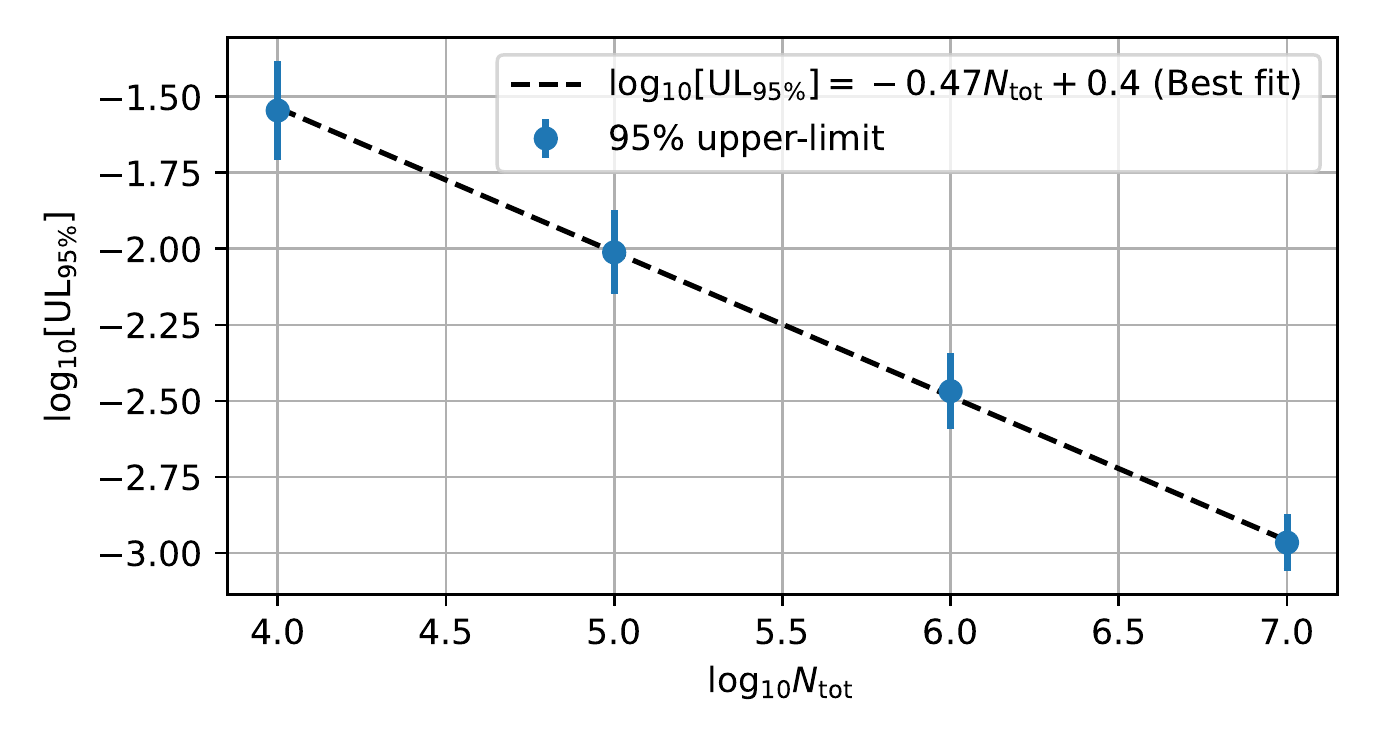}
    \caption{95\% confidence level upper-limits (blue dots) on the dipole amplitude generated from populations of BBHs isotropically distributed over the sky (no cosmic dipole present). The black dashed line indicates a fit for the upper-limit scaling. The error bars are generated by repeating each simulation 50 times.}
    \label{fig:ULtrend}
\end{figure}
To better demonstrate that the dipole detection capabilities scales as $1/\sqrt{N_{\rm tot}}$, in Fig.~\ref{fig:ULtrend}, we provide upper limits on the dipole amplitude obtained from BBHs distributions that are generated isotropically over the sky. By using $10^4, 10^5, 10^6$ and $10^7$ BBH detections simulated in the case where no cosmic dipole is present, we estimate the 95\% confidence level upper-limit on the amplitude of the cosmic dipole. The best fit for the scaling of the 95\% confidence level upper limit  $\mathrm{UL}_{95\%}$ on ${\bf \frac{v_0 \alpha}{c}}$ is given them by 
\begin{equation}
    \log_{10}[\mathrm{UL}_{95\%}]=a \log_{10}[N_{\rm tot}]+b,
\end{equation}
with $a=-0.47 \pm 0.06$ and $b= 0.4 \pm 0.4$ therefore including the typical scaling $\propto N_{\rm tot}^{-1/2}.$

From the figure, we can see that the 95\% upper limit on the dipole amplitude scales as expected with the number of detections.

In this study, we have focused on the dipole in GW number counting, showing in particular that BBHs provide a very clean way of measuring the observer velocity, since the theoretical prediction for the signal is very simple in this case. This is in contrast with the dipole from radio galaxies and quasars, for which the amplitude of the dipole depends directly on the properties of the sources, namely their spectral index and flux distribution. As shown in~\cite{Dalang:2021ruy}, if the populations evolve with redshift, then the theoretical prediction may change significantly and the tension with the CMB dipole may disappear. In this context, BBHs will provide a robust way to determine if the tension between the CMB dipole and the AGN dipole is due to our imperfect knowledge of quasars and radio galaxies properties, to systematic (that will necessarily be different with GW observations) or to a breaking of the Copernician principle.

This being said, it is still interesting to investigate if other GW estimators could help explaining the tension between the CMB and AGN dipole. In \citet{2022arXiv220407472K}, the authors propose to measure the dipole in the distribution over the sky of the detector-frame mass. Similarly to the case of number counting, detecting a cosmic dipole with this method, would require the detection of a relative discrepancy in detector-frame mass distribution of the order of $v_o/c$. To determine if this method can work, it is therefore crucial to assess the noise expected on such a measurement. Indeed, in addition to Poisson noise that affects number counting, the width of the mass distribution would induce additional contributions to the variance of the estimator. For BBHs, for which the width is expected to be large, the variance may be important~\footnote{Note that the mass distribution of BBHs is directly estimated from observed data, with complex hierarchical analysis tools such as the ones of \cite{2019ApJ...882L..24A,2021ApJ...913L...7A,2021arXiv211103634T}}. For BNSs, on the other hand, the variance may be significantly smaller since the mass distribution is much more peaked. It may also be interesting to combine mass and number counting, to determine if threshold effects can be mitigated with a specific estimator. We will explore all these aspects in a future work.

Finally, let us comment on some limitations of our prospect studies, that can be tackled in future works. In our study, we used the 0PN approximation to calculate the SNR of GW events. This is an approximation that underestimates the SNR of the detections as it neglects the merging part, which is important for BBHs. Therefore, we might expect threshold effects to be even less important than what we discussed. Another crucial assumption that we made was to neglect the sky direction dependence of the Antenna patterns of GW detectors. Due to their geometrical configuration, GW detectors are not equally sensitive to all the directions, and even for a detector network, the antenna patterns averaged during an year of observation would not be isotropic. For instance, for the simulated ET+2CE network, the antenna patterns averaged over one year of observation is 
\be
    \langle F^2\rangle\simeq -0.097 \cos^4\delta-0.040 \cos^2\delta+0.878\,,
\ee
where $\delta$ is the source declination. As ${\rm SNR}^2 \propto \langle F^2\rangle$, and $\langle F^2\rangle$ varies by about $10\%$ over the sky, we expect the SNR to vary by about $3\%$ over the sky. This means that there might be more or less sources detected over the sky due to this sensitivity variation, that may contaminate the measurement of a dipole of the order of $10^{-3}$. However, if threshold effects are negligible, as for the case of ET+2CE and BBHs, this sky-dependent selection bias will be negligible. In other cases, this effect can be calculated form first principles and used to adjust the estimator. Finally, one should consider also the sky-localization uncertainties associated to the GW detection, which could be higher than $3$ deg (although the results reported in \citet{2022arXiv220702771I} suggest this is a fair approximation for most BBH events); this would require an analysis done as in \citet{2022arXiv220705792E}. 

\section{Conclusions}
\label{sec:5}

In this paper we have discussed aspects and prospects for detecting and estimating the cosmic dipole due to the observer velocity using GWs detected with XG detectors.

In Sec.~\ref{sec:2} we have introduced the theoretical framework to evaluate the effect of the observer velocity on GW detections. We have shown that this velocity introduces an aberration on the GW localization, and that it modifies the number of detections above a given SNR threshold through the redshifted chirp mass and the luminosity distance. For BBHs, we have demonstrated that these threshold effects are negligible and that the amplitude of the GW dipole is directly given by $2 v_o/c$.

In Sec.~\ref{sec:3} we have discussed several frequentists and bayesian techniques to detect and estimate the presence of the cosmic dipole from GW counting. We have shown that, with $10^6$ BBH detections, which would be observable in a few years of observations with a ET+2CE network, it will be possible to detect a cosmic dipole with amplitude similar to the one estimated from AGN, with a precision of $\sim 16\%$. On the other hand, with $10^6$ detections, a GW dipole with amplitude compatible with that of the CMB would only be marginally detectable. With $10^7$ detections however, we would be able to significantly detect a GW dipole with amplitude compatible with both the AGN and the CMB one. If we include BNSs and model the threshold contributions to  the dipole, $10^7$ detections could be reachable in $\sim 10$ years of observations of ET+2CE.

Finally, in Sec.~\ref{sec:4} we have discussed critical aspects related to the detection of the cosmic dipole using number counting and the detector-frame mass distribution. Moreover, we have discussed the impact of some of the assumptions that we made in this exploratory study.

Next generation GW detectors, that will give us access to at least hundreds of thousands of GW detections per year, have therefore the potential to solve a growing  tension associated to the standard cosmological model.

\section*{Acknowledgements}
We thank Nicola Tamanini and Archisman Ghosh for discussions and exchanges.
S.M. is supported by the ANR COSMERGE project, grant ANR-20-CE31-001 of the French Agence Nationale de la Recherche.
S.F. is supported by the Fonds National Suisse, grant $200020\_191957$, and by the SwissMap National Center for Competence in Research. C.B. acknowledges funding from the Swiss National Science Foundation and from the European Research Council (ERC) under the European Union’s Horizon 2020 research and innovation program (Grant agreement No.~863929; project title ``Testing the law of gravity with novel large-scale structure observables"). The work of G.C. is supported by the CNRS and by Swiss National Science Foundation (Ambizione grant, ``Gravitational wave propagation in the clustered universe"). 

\section*{Data Availability}

The simulations and numerical code underlying this paper are available on GitHub \href{https://github.com/simone-mastrogiovanni/cosmic_dipole_GW_3G}{\faGithub}.

\bibliographystyle{mnras}
\bibliography{example} 

\appendix

\section{Mass population model}
\label{app:pop}

We use the \textsc{Power Law+peak} model from \citet{2021ApJ...913L...7A} to describe the source frame distribution of BBHs masses. 
This model is composed by two statistical distributions: a truncated power law distribution
\begin{equation}
\mathcal{P}(x|x_{\rm min},x_{\rm max},\alpha) \propto 
\begin{cases}
    x^\alpha & \left(x_{\rm min}\leqslant x \leqslant x_{\rm max}\right) \\
    0 & \mathrm{Otherwise}.
\end{cases}
\end{equation}
and Gaussian distribution with  mean $\mu$ and standard deviation $\sigma$.
\begin{equation}
\mathcal{G}(x|\mu,\sigma)=\frac{1}{\sigma\sqrt{2\pi}} \exp{\left[ -\frac{(x-\mu)^2}{2\sigma^2}
\right]}\,.
\end{equation}
The distribution of the source frame masses $m_{1},m_{2}$ is factorized as
\begin{equation}
\pi(m_{1},m_{2}|\Phi_m)=\pi(m_{1}|\Phi_m)\pi(m_{2}|m_{1},\Phi_m),
\end{equation}
where $\pi(m_{1}|\Phi_m)$ is the \textsc{Power Law+Peak} model and $\pi(m_{2}|m_{1},\Phi_m)$ is a truncated power law distribution. The secondary mass is conditioned to the constraint $m_{2}<m_{1}$, i.e.
\begin{equation}
    \pi(m_{2}|m_{1},m_{\rm min},\alpha)=\mathcal{P}(m_{2}|m_{\rm min},m_{1},\beta)\,.
\end{equation}
The \textsc{Power Law+Peak} distribution is
\begin{align}
    \pi(m_{1}|m_{\rm min},m_{\rm max},\alpha,\lambda_{\rm g},\mu_{\rm g},\sigma_{\rm g})=&(1-\lambda_{\rm g})\mathcal{P}(m_{1}|m_{\rm min},m_{\rm max},-\alpha)\nonumber\\&+\lambda_{\rm g} \mathcal{G}(m_{1}|\mu_{\rm g},\sigma_{\rm g})\,
    \label{eq:PLG}
\end{align}
where the power law part has slope $-\alpha$ between $m_{\rm min}$ and $m_{\rm max}$, while the gaussian component has  mean $\mu_{\rm g}$ and standard deviation $\sigma_{\rm g}$ and
accounts $\lambda_g$ total fraction of the distribution. We also apply an additional smoothing at the lower edge of the distribution
\begin{eqnarray}
\pi(m_{1},m_{2}|\Phi_m)&=&[\pi(m_{1}|\Phi_m)S(m_{1}|\delta_m,m_{\rm min})] \times  \\ &&[\pi(m_{2}|m_{1},\Phi_m)S(m_{2}|\delta_m,m_{\rm min})] \nn
\end{eqnarray}
where $S$ is a sigmoid-like window function as described in \cite{2021ApJ...913L...7A}. 

The parameters used for the simulation are $\alpha=3.4, \beta=1.1, m_{\rm max}=87 M_{\odot}, m_{\rm min}=5.1 M_{\rm \odot}, \delta_m=4.8 M_{\odot}, \sigma_g=3.6 M_{\odot}, \mu_g=34 M_{\odot}, \lambda_g=0.03$.

\bsp	
\label{lastpage}
\end{document}